# Anomalous magnetoresistance in an antiferromagnetic Kagome semimetal heterostructures

Xionghua Liu,[1,2#*] Qiyuan Feng,[3#] Weibin Cui,[4#] Hanjie Guo,[5] Yubin Hou,[3] Xiaomin Zhang,[1,2] Yongcheng Deng,[1,2] Dong Zhang,[1,2] Jing Zhang,[1,2] Qingyou Lu,[3*] Kaiyou Wang[1,2*]

[1] *State Key Laboratory of Semiconductor Physics and Chip Technologies, Institute of Semiconductors, Chinese Academy of Sciences, Beijing 100083, China*
[2] *Center of Materials Science and Optoelectronics Engineering, University of Chinese Academy of Sciences, Beijing 100049, China*
[3] *Anhui Province Key Laboratory of Condensed Matter Physics at Extreme Conditions, High Magnetic Field Laboratory, Chinese Academy of Sciences, Hefei, Anhui, China.*
[4] *Key Laboratory of Electromagnetic Processing of Materials Ministry of Education, Northeastern University, Shenyang 110819, China*
[5] *Songshan Lake Materials Laboratory, Dongguan, Guangdong 523808, China*

**Antiferromagnetic Kagome semimetals have attracted tremendous attentions for their potential application in antiferromagnetic topological spintronics. Effectively manipulating Kagome antiferromagnetic states could reveal abundant physical phenomena induced from quantum interactions between topology, spin, and correlation. Here, we achieved tunable spin textures of FeSn thin films via introducing interfacial Dzyaloshinskii – Moriya interaction from heavy-metal Pt overlayer. With increasing FeSn thickness, the variable spin textures result in gradual change in Hall resistivity and magnetoresistance. Importantly, an unconventional damped oscillatory-like behavior of magnetoresistance at relatively low magnetic field can be observed in thin FeSn/Pt samples. This oscillatory-like magnetoresistance feature was confirmed to be related to the special topological spin textures revealed by magnetic force microscopy measurements. The formation of rich variety of topological spin textures in association with exotic magneto-transport properties in antiferromagnetic Kagome FeSn heterostructures offers new perspectives for understanding the novel emergent phenomena in Kagome antiferromagnets.**

# 1. Introduction

The Kagome lattice is composed of hexagons and triangles in a two-dimensional (2D) network of corner-shared triangles that possesses geometric frustration, nontrivial band topology, van Hove singularities, and flat bands,[1,2] hence it is a versatile platform to investigate the interplay of various degrees of freedom such as charge, spin, orbital, and lattice.[3,4] Recently, diverse interesting phenomena have been discovered in Kagome materials, such as unconventional superconductor,[5,6] quantum spin liquid,[7] anomalous Hall effect,[8] and topological semimetals.[9,10] Moreover, antiferromagnets have attracted extensive interests in spintronics as one of the active materials for next-generation spintronic devices, with the prospect of supplying high density memory integration and ultrafast data processing.[11-15] Therefore, manipulating spin states in antiferromagnetic (AFM) Kagome materials and studying their magnetotransport behaviors have become a fascinating issue in the condensed matter physics and information technology communities.

Compared to the previously studied AFM Weyl semimetal $Mn_3Sn$,[16-18] FeSn consists of an alternating stack of 2D Fe Kagome layers and 2D Sn honey-comb (stanene) layers, which provides an ideal playground to investigate the physical properties of the 2D Kagome network in a bulk crystal. FeSn is magnetically ordered, having Fe moments ferromagnetically aligned within each Kagome plane but antiferromagnetically coupled along the c-axis (**Figure 1(a)**) with Néel temperature $T_N \approx 368$ K.[19] This magnetic structure permits a simple hopping model rather than the complicated non-collinear spin textures observed in $Mn_3Sn$-type Kagome antiferromagnets. The angular resolved photoemission spectroscopy (ARPES) experiment has revealed the Dirac fermions and flat bands in FeSn.[10,20,21] To further explore the interaction between band topology and magnetism and develop the application of FeSn in topological spintronics, an important step is to effectively manipulate its spin textures and investigate the corresponding transport properties.

The Dzyaloshinskii – Moriya interaction (DMI),[22,23] induced by strong spin-orbit coupling and broken inversion symmetry, offers an effective means to modify a relative tilt between neighboring spins. Here we designed and achieved various spin states in antiferromagnetic Kagome FeSn capped with a Pt overlayer system via interfacial DMI, which presents exotic phenomena in magneto-transport properties. The highly tunable spin textures in an antiferromagnetic Kagome semimetal would enlarge the designability of topological antiferromagnetic spintronics.

## 2. Results and Discussion

### 2.1 Microstructural characterizations

To vary the spin textures of FeSn thin film, we introduced the interfacial DMI with a Pt overlayer because the spin textures of AFM Kagome $Mn_3Sn$ can be gradually reconstructed from co-planar inverted triangular structures to Bloch-type skyrmions via tuning the magnitude of interfacial DMI in our previous work.[24,25] The experiments were performed on $SrTiO_3$ (111)/FeSn ($t_{FeSn}$)/Pt (3) (thickness in nanometers) thin films with $t_{FeSn}$ = 2.7, 6.6, 10, 20 and 30 nm. The Pt overlayer was kept at 3 nm due to the nearly saturated interfacial DMI for this thickness [24]. Here the FeSn and Pt thin films were grown by MBE and magnetron sputtering, respectively (see Experimental Section for details). The microstructural characterizations of our FeSn thin films were first conducted using *in situ* RHEED patterns, the sharp RHEED streaks indicate a flat and well-ordered single-crystalline surface structure of the FeSn films (**Figure 1(b)**). Moreover, the XRD patterns do not display any phase other than substrate and FeSn (002) (**Figure 1(c)**). Besides, the high-resolution transmission electron microscope (HR-TEM) shows two types of layers stacked alternatively, corresponding to the $Fe_3Sn$ and $Sn_2$ layers (as the blue and red arrows indicate), respectively, and the atomic structure is the same as FeSn (**Figure 1(d)**). These characterizations confirm high quality of our FeSn thin films.

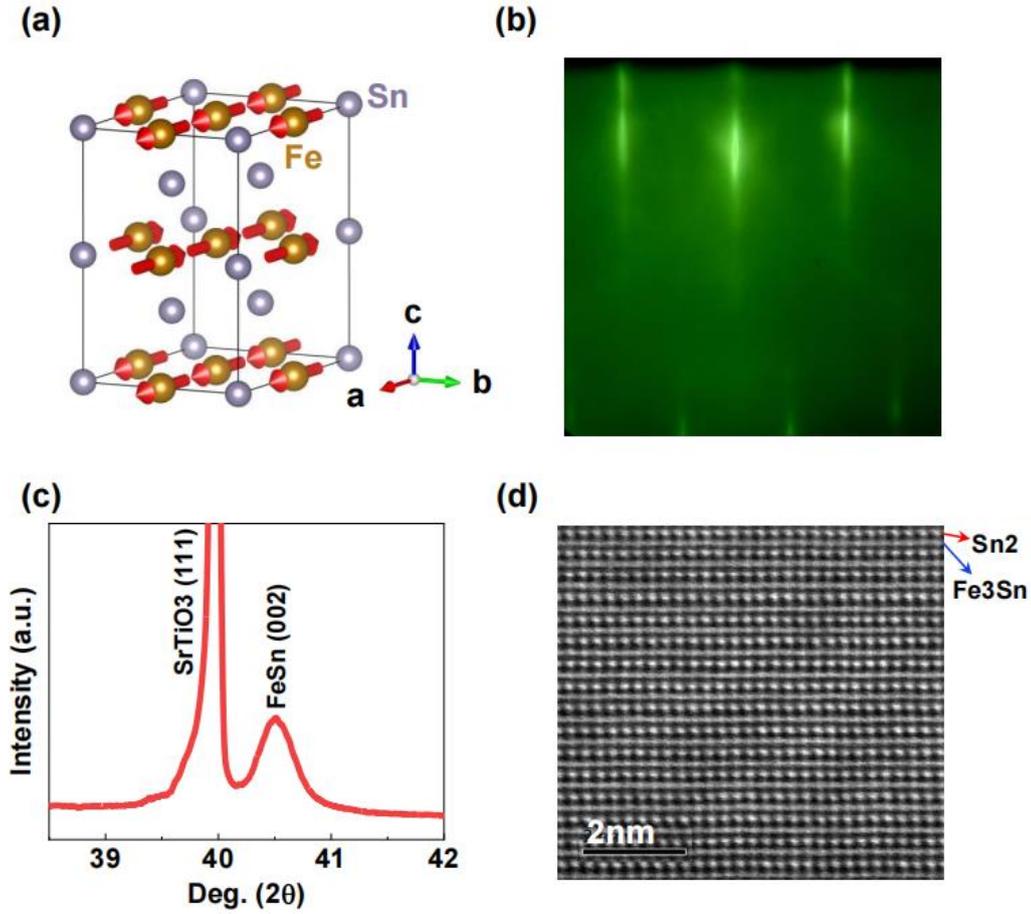

**Figure 1. Characterizations of FeSn thin film. a)** FeSn spin/lattice structure. It consists of an alternating stack of 2D Fe Kagome layers and 2D Sn honey-comb (stanene) layers with antiferromagnetically coupled along the c-axis. **b)** RHEED patterns of FeSn thin film, the sharp RHEED strips suggest a flat and well-ordered single-crystalline surface structure of the film. **c)** XRD pattern for FeSn thin film, it displays the FeSn (002) crystal orientation. **d)** HR-TEM image of the cross section of FeSn film, it shows $Fe_3Sn$ layer and $Sn_2$ stacked alternatively (as the arrows indicate).

## 2.2 Hall resistivity

Subsequently, we focused on the magneto-transport properties of our FeSn ($t_{FeSn}$)/Pt (3 nm) heterostructures, see the measurement scheme and fabricated device in Supplementary **Figure S1**. We first performed the Hall resistivity as a function of the out-of-plane magnetic field $\rho_H$ *vs.* $B$ curves with varying thickness of FeSn samples at room temperature (see Supplementary **Figure S2**). The evolution of $\rho_H$ with $t_{FeSn}$ can be observed, that the anomalous Hall resistivity $\rho_{AHE} = \rho_H - \rho_{OHE}$ gradually changes from negative to positive under 1.5 T (see Supplementary **Figure S3**), here $\rho_{OHE} = R_0 B$, $R_0$ and $B$ are the ordinary Hall coefficient and out-of-plane magnetic field, respectively. Interestingly, the topological Hall effect (THE) was noticed for FeSn (10 nm)/Pt (3 nm) sample, this means the variable magnetic structures and even the topological spin textures occur in our FeSn ($t_{FeSn}$)/Pt (3 nm) due to the strong interfacial DMI.[24-26]

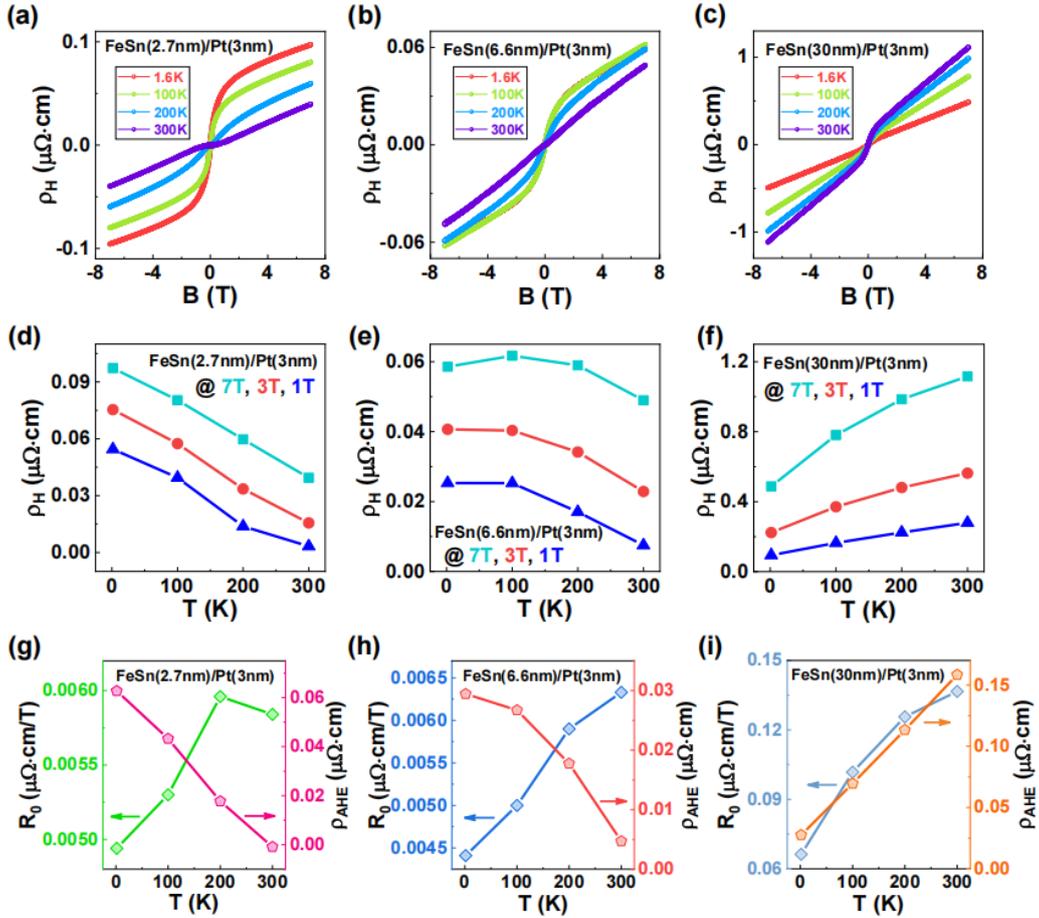

**Figure 2. Temperature dependence of Hall resistivity in FeSn/Pt heterostructures. a-c)** $\rho_H$ *vs.* $B$ curves measured at 1.6, 100, 200 and 300 K for FeSn ($t_{FeSn}$)/Pt (3 nm) samples with $t_{FeSn}$ = 2.7 nm (**a**), 6.6 nm (**b**) and 30 nm (**c**). **d-f)** $\rho_H$ *vs.* $T$ curves under 7 T, 3 T and 1 T for $t_{FeSn}$ = 2.7 nm (**d**), 6.6 nm (**e**) and 30 nm (**f**). With increasing temperature, the $\rho_H$ gradually decreases and increases for thin (2.7 nm) and thick (30 nm), respectively, while the $\rho_H$ keeps nearly constant at low temperature and slightly reduces at high temperature for moderate one (6.6 nm). **g-i)** Temperature dependent on ordinary Hall coefficient $R_0$ and anomalous Hall resistivity $\rho_{AHE}$ for $t_{FeSn}$ = 2.7 nm

(**g**), 6.6 nm (**h**) and 30 nm (**i**). With the temperature increases, $\rho_{AHE}$ gradually reduces for $t_{FeSn}$ = 2.7 and 6.6 nm but enhances for $t_{FeSn}$ = 30 nm, this opposite temperature dependent evolution of AHE for thin and thick FeSn samples were linked to the variable spin textures.

We chose three typical samples (FeSn ($t_{FeSn}$)/Pt (3 nm) with $t_{FeSn}$ = 2.7, 6.6 and 30 nm) to further investigate the temperature dependent Hall characters. Clear variation of $\rho_H$ *vs.* B curves can be observed for three samples in **Figure 2(a)-(c)**. With increasing temperature, the $\rho_H$ gradually decreases for 2.7-nm-thick FeSn, while slightly reduces for 6.6-nm-thick FeSn, but enhances for 30-nm-thick FeSn samples, respectively (**Figure 2(d)-(f)**). The opposite trend of $\rho_H$ with temperature for thin (2.7 nm) and thick (30 nm) FeSn samples means different spin textures, and the later one has comparable properties to the pure FeSn film.[27,28] Remarkably, the FeSn (6.6 nm)/Pt (3 nm) sample presents nearly constant $\rho_H$ at low temperature (**Figure 2(e)**) accompanying with very small $\rho_{AHE}$ signal (see **Figure 2(g)-(i)**), suggesting quite weak contribution from the magnetic moment for this sample.[14,29,30] In addition, the $R_0$ and $\rho_{AHE}$ of the thinner FeSn/Pt samples (see **Figure 2(g)-(h)**) exhibit about one order of magnitude smaller than that of the thick one (**Figure 2(i)**) as well as the bulk FeSn.[31] This indicates the spin textures of FeSn have been rebuilt by interfacial DMI for our thin FeSn/Pt samples.[22-26]

Similar to the Mn$_3$Sn/Pt heterostructures, two dominated competition interactions should exist for our FeSn ($t_{FeSn}$)/Pt (3 nm) samples, including the interfacial DMI between outer few Fe$_3$Sn sublayer and Pt layer and the effective interlayer antiferromagnetic exchange interaction between Kagome sublayers. The thicker the FeSn film the weaker effect from the interfacial DMI on spin textures. As the effective interlayer antiferromagnetic exchange interaction becomes stronger with lowering temperature because of the smaller thermal effect, therefore, the competition between above mentioned two interactions result in different temperature dependent

Hall resistivity in our FeSn ($t_{FeSn}$)/Pt (3 nm) heterostructures.[22-26]

## 2.3 Anomalous magnetoresistance

We next came to the investigation on temperature dependence of magnetoresistance ($MR = [\rho(B) − \rho(0)]/\rho(0) \times 100\%$) of our FeSn ($t_{FeSn}$)/Pt (3 nm) samples. For moderate thick FeSn sample (FeSn (6.6 nm)/Pt (3 nm)), the $MR$ vs. $B$ curve displays parabolic behavior under high magnetic field ($B > 4$ T) both for in-plane and out-of-plane conditions at very low temperature 1.6 K (see **Figure 3(a)-(b)**). Differing from the out-of-plane negative $MR$ at low field, the in-plane positive $MR$ shows a crossover from increasing to decreasing value below 2 T (**Figure 3(a)**). This positive $MR$ at low magnetic field ($B < 0.4$ T) would be induced from weak-antilocalization like that found in some topological materials.[32-34] We then tried to understand the gradual reduction of $MR$ between 0.4 and 2 T. It has been reported that for a weak Kondo effect system, the negative $MR$ significantly deviates from the quadratic dependence and tends to be saturated at high magnetic fields.[35] Consequently, the $MR$ from the Lorentz force could be extracted by quadratically fitting the $MR$ behavior in the high-field region (from 4 to 7 T), see the blue dashed lines in **Figure 3(a)** and **3(b)**. Then the experimental results were subtracted by the fitted data. The obtained $\Delta MR\%$ = $MR\%_{Exp}$ - $MR\%_{Fit}$ curves would represent contributions from the Kondo effect, see Supplementary **Figure S4**. We therefore measured the resistance with temperature ($R_{xx}$ vs. $T$ curve) for our FeSn (6.6 nm)/Pt (3 nm) sample to confirm it. As exhibited in Supplementary **Figure S5** the minimum resistance can be observed at 9 K, hence the Kondo effect should take important role in $MR$ at 1.6 K in our system (**Figure 3(a)-(b)**).[36,37]

Important is that the $MR$ against $B$ curve at high temperature (>50 K) displays an oscillation-like property (see **Figure 3(c)-(d)**). Such as the data at 100 K (**Figure 3(c)**), one notes the $MR$ exhibits an initial increase to a maximum at ~0.65 T and then reduces to a minimum at ~1.7 T. The second maximum and minimum values can be observed at ~3.1 T and 4.9 T, respectively, but the amplitude becomes decaying with magnetic fields. Notably, this oscillatory-like behavior differs completely from those of the Shubnikov – de Haas oscillations in that it exhibits a damped like-periodicity as a function of $B$ rather than of $1/B$.[38-40] The later phenomenon is usually found in topological materials,[41,42] two-dimensional electron gas system,[43] superconducting films[44] or 2D antiferromagnet.[45] This low-field $MR$ oscillation-like character was rarely reported, but it can be observed for some special system possessing strong competition between bulk and surface $MR$ when the bulk $MR$ should not be so large as to overwhelm the surface $MR$.[46,47] In Sondheimer's theory,[48] the variation of $\sigma_B/\sigma$ is plotted versus a dimensionless parameter $\gamma = a/r = \frac{aeB}{m^*\bar{v}c}$ for fixed values of another dimensionless parameter $\kappa = a/l_B$ and of the surface reflection coefficient ($p$). The $\sigma_B$ and $\sigma$ are the bulk and the total conductivity, $a$ is the thickness of the film, $l_B$ is the bulk mean free path, $m^*$ is the effective mass, $\bar{v}$ is the Fermi velocity. The coefficient $p$ ranges from $p = 0$ for completely diffuse scattering of the electrons at the sample boundaries to $p = 1$ for completely specular reflection. For small $\kappa$ and $p$, the relatively large $\sigma_B/\sigma$ would present a damped oscillatory behavior with $B$,[48] thus this phenomenon could be seen in the sample with suitable $\sigma_B/\sigma$, $a/l_B$, and $p$.

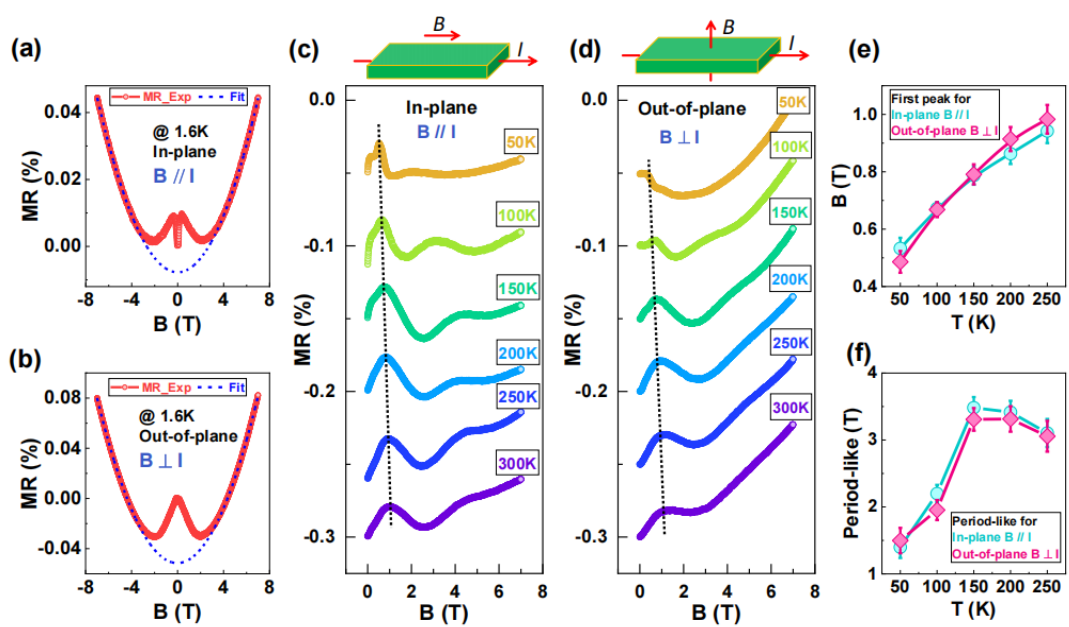

**Figure 3. Magnetoresistance with temperature for FeSn (6.6 nm)/Pt (3 nm) sample. a-b)** The *MR vs. B* curves at 1.6 K for magnetic field parallel (**a**) and perpendicular (**b**) to film plane and electrical current. The *MR* under high magnetic field can be well quadratically fitted with blue dashed curves. The small positive *MR* under low magnetic field for in-plane (**a**) would be induced from the weak-antilocalization, while the negative *MR* at moderate field in (**a**) and at $B < 4$ T in (**b**) may be related to the Kondo effect. **c-d)** The *MR vs. B* curves at 50, 100, 150, 200, 250 and 300 K for in-plane (**c**) and out-of-plane (**d**), respectively. The damped oscillatory-like behavior in *MR* can be observed. **e)** Magnetic field corresponding to the first maximum *MR* (as indicated with dashed lines in (**c**) and (**d**)) versus temperature for in-plane and out-of-plane. **f)** Period-like field defined as the length between the first and second maxima *MR*. A rapid increase in period-like field can be observed round 100 K.

In our FeSn ($t_{FeSn}$)/Pt (3 nm) samples, the moderate FeSn (FeSn (6.6 nm)/Pt (3 nm)) sample probably satisfies above conditions, that the $a/l_B$ and $p$ could be small and the $\sigma_B/\sigma$ keeps relatively large. For very thin FeSn sample (FeSn (2.7 nm)/Pt (3 nm)), the negative *MR* for out-of-plane condition can be observed at low temperature (Supplementary **Figure S6**), indicating the spin direction of the Fe in FeSn would be partially rotated to out-of-plane owing to the strong interfacial DMI;[24-26] for thicker FeSn samples, the FeSn layer mostly maintains its nature spin textures. As a result, the *MR* vs. *B* curves of the thinner (FeSn (2.7 nm)/Pt (3 nm)) or thicker (FeSn (10 nm)/Pt (3 nm)) FeSn samples display weak oscillatory-like behavior, see Supplementary **Figure S7**.

Furthermore, one notes the magnetic field corresponding to the first maximum *MR* gradually shifts to higher magnetic field (see the dashed lines in **Figure 3(c)-(d)**) ranging from around 0.5 to 1 T (**Figure 3(e)**). We defined the period-like *MR* of our system as the length between the first and second maxima *MR*, a rapid increase in period-like field at around 100 K can be seen (**Figure 3(f)**). For the system having damped oscillatory-like property of magnetoresistance, the first observed maximum was reported to occur at $\omega\tau \approx 1$, here $\omega = eB/m^*c$ is the cyclotron resonance frequency, $\tau$ represents the collision relaxation time.[46-48] The $\tau$ becomes shorter at higher temperature, accordingly the $\omega$ should get larger with the higher magnetic field, consistent with our result in **Figure 3(e)**. The different period-like *MR* with temperature would be related to different types of spin textures.

## 2.4 Magnetic force microscopy measurements

To verify the effect of spin textures on the period-like *MR* in our system, we employed real-space magnetic force microscopy (MFM) to examine the magnetic structures of our FeSn (6.6 nm)/Pt (3 nm) heterostructures. To effectively separate these magnetic structures, we applied pixel-by-pixel subtraction operations to adjacent MFM images.[49-51] **Figure 4(a)** shows the evolution of MFM images with enhancing magnetic field from 0 to 8 T measured at 5 K. One notes the large area magnetic structure coherently reverses at moderate magnetic field, which is similar to the $Mn_3Sn$/Pt heterostructures at low temperature because of the probably formation of antiferromagnetic meron-like spin textures in $Mn_3Sn$/Pt sample.[24,52,53] For comparison, at high temperature 110 K, the MFM images display complex mixed states including skyrmions and conventional magnetic domains, see 1, 2, 3 positions under different magnetic fields in **Figure 4(b)**. The damped oscillatory-like *MR* phenomenon cannot be found at low temperature (**Figure 3(a)-(b)**) but appears at relatively high temperature (**Figure 3(c)-(d)**). Therefore, clear temperature dependence of spin textures in our FeSn (6.6 nm)/Pt (3 nm) sample could be responsible for the variation of damped oscillatory-like *MR* in **Figure 3(f)**, but the exact relationship between the spin textures and period-like *MR* character needs to be further investigated.

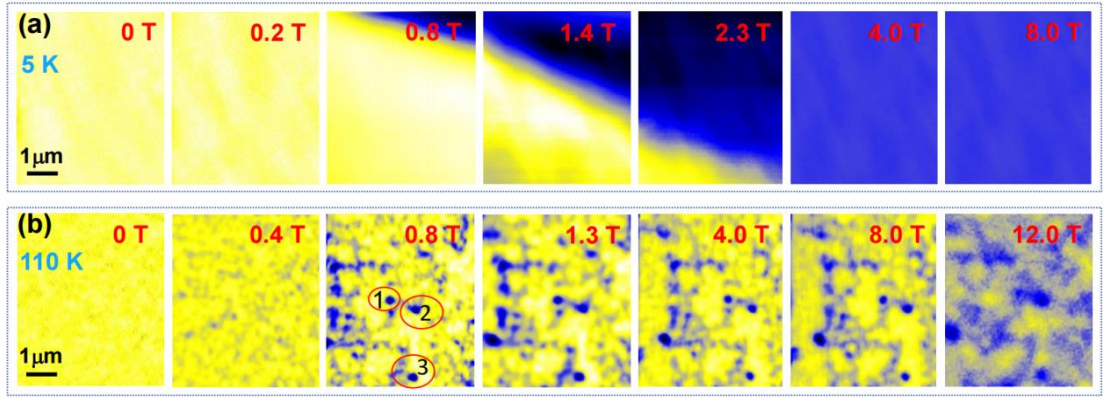

**Figure 4 Magnetic force microscopy images for FeSn (6.6 nm)/Pt (3 nm) sample.** **(a)** MFM patterns measured at 5 K with enhancing magnetic field from 0 to 8 T. A coherent reversal of magnetic structure can be observed for moderate magnetic field. **(b)** MFM patterns measured at 110 K with the magnetic field increases from 0 to 12 T. A complex mixed states including skyrmions and conventional magnetic domains can be observed, see the skyrmions structure in 1, 2, 3 positions (indicated with red circles). This different magnetic structure results in quite different magneto-transport properties.

## 3. Conclusion

In summary, we have investigated the temperature dependent on magneto-transport properties in FeSn ($t_{FeSn}$)/Pt (3 nm) heterostructures with $t_{FeSn}$ = 2.7, 6.6, 10, 20 and 30 nm. With increasing FeSn thickness, the evolution of spin textures from skyrmions to pure FeSn can be observed because of the interfacial DMI. Correspondingly, the Hall resistivity and *MR* gradually vary with the FeSn thickness. Significantly, unlike the normal observed *MR* oscillatory under very high magnetic field, a damped oscillatory-like behavior of *MR* at relatively low magnetic field can be achieved in our thin FeSn thick samples. The MFM measurements revealed the variable magnetic structures of FeSn/Pt sample at different temperature, the oscillatory-like *MR* feature could be greatly dependent on the special topological spin textures (skyrmions) owing to the strong competition between bulk and surface *MR*. Our work demonstrates that the rich variety of topological spin textures could be formed in antiferromagnetic Kagome FeSn heterostructures with abundant magneto-transport properties, which may promote the discovery of novel emergent phenomena in Kagome antiferromagnets.

## 4. Experimental Section

The samples used for investigating the magnetotransport properties in AFM Kagome semimetal heterostructures consist of FeSn ($t_{FeSn}$)/Pt (3) (thickness in nanometers) thin films with $t_{FeSn}$ = 2.7, 6.6, 10, 20 and 30 nm grown on SrTiO$_3$ (111) substrates. The FeSn thin films were grown by molecular beam epitaxy (MBE) in an ultrahigh vacuum (UHV) chamber with a base pressure in the $5 \times 10^{-11}$ Torr range. High purity Fe metal and Sn metal were evaporated from LUXEL Radak effusion cells at temperatures of about 1290°C and 990°C, respectively. The substrates were annealed for 1 h at 600°C in an oxygen pressure of $1.2 \times 10^{-2}$ Torr to obtain a clean and well-ordered surface structure prior to the FeSn deposition. The substrate temperature was kept at 500°C during growth, and the Fe and Sn fluxes were calibrated using a quartz-crystal monitor (QCM) at the growth position prior to deposition and set to 0.7 Å and 1.1 Å per minute, respectively, for the growth of all films. After growing FeSn thin films, a 10 nm-thick Te thin film was grown at room temperature by MBE as protecting layer. Then the FeSn/Te films were moved to magnetron sputtering chamber with a base pressure in the $1 \times 10^{-8}$ Torr range. The FeSn/Te films were heated to 300°C to remove the Te film, and then the 3 nm-thick Pt layer was grown by magnetron sputtering at room temperature.

*In situ* and *real-time* monitoring of the epitaxial growth was performed by reflection high-energy electron diffraction (RHEED) measurements. X-ray diffraction (XRD) was employed for further *ex situ* investigation of the structural quality and the microstructure of the films. The sample for cross-sectional scanning transmission electron microscopy characterization was prepared using a Zeiss Auriga focused ion beam system. The high-angle annular dark-field (HAADF) Z-contrast images, bright-field images were acquired using a spherical aberration-corrected FEI Titan

Cubed Themis 60-300 operated at 200 kV. The samples were patterned into Hall bar devices with a current channel width of 10 μm using the standard photolithography and Ar-ion etching (see the device in Supplementary **Figure S1**). A home-built variable temperature magnetic force microscopy (MFM) equipped with a 20 T commercial superconducting magnet (Oxford Instruments) was used. For FeSn/Pt heterostructures, there is negligible force (attractive force) between the local moments of antiferromagnetic phase and the MFM tip, which cause a bright contrast (dark contrast) in the MFM images. MFM images were analyzed using Gwyddion software.

# References

[1] J. X. Yin, B. Lian, and M. Z. Hasan, *Nature* **2022**, 612, 647.

[2] Y. J. Wang, H. Wu, G. T. McCandless, J. Y. Chan, and M. N. Ali, *Nat. Rev. Phys*. **2023**, 5, 635.

[3] Q. Wang, H. C. Lei, Y. P. Qi, and C. Felser, *Acc. Mater. Res*. **2024**, 5, 786.

[4] H. R. Zhang, H. F. Feng, X. Xu, W. Hao, and Y. Du, *Adv. Quantum Technol*. **2021**, 4, 2100073.

[5] B. R. Ortiz, L. C. Gomes, J. R. Morey, M. Winiarski, M. Bordelon, J. S. Mangum, I. W. H. Oswald, J. A. Rodriguez-Rivera, J. R. Neilson, S. D. Wilson, E. Ertekin, T. M. McQueen, and E. S. Toberer, *Phys. Rev. Mater*. **2019**, 3, 094407.

[6] B. R. Ortiz, S. M. L. Teicher, Y. Hu, J. L. Zuo, P. M. Sarte, E. C. Schueller, A. M. M. Abeykoon, M. J. Krogstad, S. Rosenkranz, R. Osborn, R. Seshadri, L. Balents, J. He, and S. D. Wilson, *Phys. Rev. Lett*. **2020**, 125, 247002.

[7] M. R. Norman, *Rev. Mod. Phys*. **2016**, 88, 041002.

[8] S. Nakatsuji, N. Kiyohara, and T. Higo, *Nature* **2015**, 527, 212.

[9] L. Ye, M. Kang, J. Liu, F. von Cube, C. R. Wicker, T. Suzuki, C. Jozwiak, A. Bostwick, E. Rotenberg, D. C. Bell, L. Fu, R. Comin, and J. G. Checkelsky, *Nature* **2018**, 555, 638.

[10] M. Kang, L. Ye, S. Fang, J. You, A. Levitan, M. Han, J. I. Facio, C. Jozwiak, A. Bostwick, E. Rotenberg, M. K. Chan, R. D. McDonald, D. Graf, K. Kaznatcheev, E. Vescovo, D. C. Bell, E. Kaxiras, J. Brink, M. Richter, M. P. Ghimire, J. G. Checkelsky, and R. Comin, *Nat. Mater*. **2020**, 19, 163.

[11] T. Jungwirth, X. Marti, P. Wadley, and J. Wunderlich, *Nat. Nanotech*. **2016**, 11, 231.

[12] V. Baltz, A. Manchon, M. Tsoi, T. Moriyama, T. Ono, and Y. Tserkovnyak, *Rev. Mod. Phys*. **2018**, 90, 015005.

[13] L. Šmejkal, Y. Mokrousov, B. H. Yan, and A. H. MacDonald, *Nat. Phys*. **2018**, 14, 242.

[14] L. Šmejkal, A. H. MacDonald, J. Sinova, S. Nakatsuji, and T. Jungwirth, *Nat. Rev. Mater.* **2022**, 7, 482.

[15] X. H. Liu, K. W. Edmonds, Z. P. Zhou, and K. Y. Wang, *Phys. Rev. Applied* **2020**, 13, 014059.

[16] H. Tsai, T. Higo, K. Kondou, T. Nomoto, A. Sakai, A. Kobayashi, T. Nakano, K. Yakushiji, R. Arita, S. Miwa, Y. Otani, and S. Nakatsuji, *Nature* **2020**, 580, 608.

[17] Y. Takeuchi, Y. Yamane, J. Yoon, R. Itoh, B. Jinnai, S. Kanai, J. Ieda, S. Fukami, and H. Ohno, *Nat. Mater.* **2021**, 20, 1364.

[18] Y. C. Deng, X. H. Liu, Y. Y. Chen, Z. Z. Du, N. Jiang, C. Shen, E. Z. Zhang, H. Z. Zheng, H. Z. Lu, and K. Y. Wang, *National Science Review* **2023**, 10, nwac154.

[19] L. Häggström, T. Ericsson, R. Wäppling, and K. Chandra, *Phys. Scr.* **1975**, 11, 47.

[20] Z. Lin, C. Wang, P. Wang, S. Yi, L. Li, Q. Zhang, Y. Wang, Z. Wang, H. Huang, Y. Sun, Y. Huang, D. Shen, D. Feng, Z. Sun, J. Cho, C. Zeng, and Z. Y. Zhang, *Phys. Rev. B* **2020**, 102, 155103.

[21] M. Han, H. Inoue, S. Fang, C. John, L. Ye, M. K. Chan, D. Graf, T. Suzuki, M. P. Ghimire, W. J. Cho, E. Kaxiras, and J. G. Checkelsky, *Nat. Commun.* **2021**, 12, 5345.

[22] I. Dzyaloshinsky, *J. Phys. Chem. Solids* **1958**, 4, 241.

[23] T. Moriya, *Phys. Rev.* **1960**, 120, 91.

[24] X. H. Liu, Q. Y. Feng, D. Zhang, Y. C. Deng, S. Dong, E. Z. Zhang, W. H. Li, Q. Y. Lu, K. Chang, and K. Y. Wang, *Adv. Mater.* **2023**, 35, 2211634.

[25] X. H. Liu, D. Zhang, Y. C. Deng, N. Jiang, E. Z. Zhang, C. Shen, K. Chang, and K. Y. Wang, *ACS Nano* **2024**, 18, 1013.

[26] H. X. Yang, J. H. Liang, Q. R. Cui, *Nat. Rev. Phys.* **2023**, 5, 43.

[27] H. Inoue, M. Han, L. Ye, T. Suzuki, and J. G. Checkelsky, *Appl. Phys. Lett.* **2019**, 115, 072403.

[28] D. Hong, C. Liu, H. Hsiao, D. Jin, J. E. Pearson, J. M. Zuo, and A. Bhattacharya, *AIP Advances* **2020**, 10, 105017.

[29] N. Nagaosa, J. Sinova, S. Onoda, A. H. MacDonald, and N. P. Ong, *Rev. Mod. Phys.* **2010**, 82, 1539.

[30] A. Bandyopadhyay, N. B. Joseph, and A. Narayan, *Materials Today Electronics* **2024**, 8, 100101.

[31] B. C. Sales, J. Yan, W. R. Meier, A. D. Christianson, S. Okamoto, and M. A. McGuire, *Phys. Rev. Mater.* **2019**, 3, 114203.

[32] E. McCann, K. Kechedzhi, V. I. Falko, H. Suzuura, T. Ando, and B. L. Altshule, *Phys. Rev. Lett.* **2006**, 97, 146805.

[33] S. Thomas, D. J. Kim, S. B. Chung, T. Grant, Z. Fisk, and J. Xia, *Phys. Rev. B* **2016**, 94, 205114.

[34] H. Z. Lu, and S. Q. Shen, *Chin. Phys. B* **2016**, 25, 117202.

[35] Y. Wang, C. Xie, J. Li, Z. Du, L. Cao, Y. Han, L. Zu, H. Zhang, H. Zhu, X. Zhang, Y. Xiong, and W. S. Zhao, *Phys. Rev. B* **2021**, 103, 174418.

[36] H. Liu, Y. Xue, J. A. Shi, R. A. Guzman, P. Zhang, Z. Zhou, Y. He, C. Bian, L. Wu, R. Ma, J. Chen, J. Yan, H. Yang, C. Shen, W. Zhou, L. Bao, and H. J. Gao, *Nano Lett*. **2019**, 19, 8572.

[37] J. Ge, T. Luo, Z. Lin, J. Shi, Y. Liu, P. Wang, Y. Zhang, W. Duan, and J. Wang, *Adv. Mater*. **2021**, 33, 2005465.

[38] J. Hajdu, *Modern Problems in Condensed Matter Sciences* **1991**, 27, 997.

[39] D. G. Seiler, and A. E. Stephens, *Modern Problems in Condensed Matter Sciences* **1991**, 27, 1031.

[40] B. Fallahazad, H. C. P. Movva, K. Kim, S. Larentis, T. Taniguchi, K. Watanabe, S. K. Banerjee, and E. Tutuc, *Phys. Rev. Lett*. **2016**, 116, 086601.

[41] H. Wang, H. Liu, Y. Li, Y. Liu, J. Wang, J. Liu, J. Dai, Y. Wang, L. Li, J. Yan, D. Mandrus, X. C. Xie, and J. Wang, *Sci. Adv*. **2018**, 4, eaau5096.

[42] S. Barua, M. R. Lees, G. Balakrishnan, and P. A. Goddard, *Phys. Rev. B* **2024**, 110, 155113.

[43] A. D. Caviglia, S. Gariglio, C. Cancellieri, B. Sacepe, A. Fete, N. Reyren, M. Gabay, A. F. Morpurgo, and J. M. Triscone, *Phys. Rev. Lett*. **2010**, 105, 236802.

[44] I. Sochnikov, A. Shaulov, Y. Yeshurun, G. Logvenov, and I. Bozovic, *Nat. Nanotech*. **2010**, 5, 516.

[45] P. Shi, X. Wang, L. Zhang, W. Song, K. Yang, S. Wang, R. Zhang, L. Zhang, T. Taniguchi, K. Watanabe, S. Yang, L. Zhang, L. Wang, W. Shi, J. Pan, and Z. Wang, *Phys. Rev. X* **2024**, 14, 041065.

[46] E. H. Sondheimer, *Nature* **1949**, 164, 920.

[47] J. Babiskin, and P. G. Siebenmann, *Phys. Rev*. **1957**, 107, 1249.

[48] E. H. Sondheimer, *Phys. Rev*. **1950**, 80, 401.

[49] L. Wang, Q. Feng, Y. Kim, R. Kim, K. H. Lee, S. D. Pollard, Y. J. Shin, H. Zhou, W. Peng, D. Lee, W. Meng, H. Yang, J. H. Han, M. Kim, Q. Lu, and T. W. Noh, *Nat. Mater*. **2018**, 17, 1087.

[50] W. B. Wang, Y. Ou, C. Liu, Y. Y. Wang, K. He, Q. K. Xue, and W. D. Wu, *Nat. Phys*. **2018**, 14, 791.

[51] H. B. Zhou, Z. Wang, Y. B. Hou, and Q. Y. Lu, *Ultramicroscopy* **2014**, 147, 133.

[52] H. Jani, J. Lin, J. Chen, J. Harrison, F. Maccherozzi, J. Schad, S. Prakash, C. B. Eom, A. Ariando, T. Venkatesan, and P. G. Radaelli, *Nature* **2021**, 590, 74.

[53] Y. Q. Zhou, S. Li, X. Liang, and Y. Zhou, *Adv. Mater*. **2025**, 37, 2312935.